\documentclass[useAMS,usenatbib]{mn2e}
\usepackage{natbib}
\usepackage{epsfig}
\usepackage{amsbsy}
\usepackage{hyperref}

\providecommand{\adsurl}[1]{\href{#1}{ADS}}

\newcommand{\lya}{Lyman-$\alpha$~}

\newcommand{\be}{\begin{equation}}
\newcommand{\ee}{\end{equation}}
\newcommand{\ba}{\begin{eqnarray}}
\newcommand{\ea}{\end{eqnarray}}
\newcommand{\brr}{\begin{array}}

\newcommand{\err}{\end{array}}
\newcommand{\bc}{\begin{center}}
\newcommand{\ec}{\end{center}}

\DeclareMathAlphabet{\mathsc}{OT1}{cmr}{m}{sc}
\def\testbx{bx}%
\DeclareRobustCommand{\ion}[2]{%
\relax\ifmmode
\ifx\testbx\f@series
{\mathbf{#1\,\mathsc{#2}}}\else
{\mathrm{#1\,\mathsc{#2}}}\fi
\else\textup{#1\,{\mdseries\textsc{#2}}}%
\fi}
\title[Primordial non-Gaussianities in the Intergalactic Medium]
{Primordial non-Gaussianities in the Intergalactic Medium}

\author[M. Viel et al.] 
{M.  Viel$^{1,2}$
E. Branchini$^{3}$,
K. Dolag$^{4}$,
M. Grossi$^4$,
S. Matarrese$^{5,6}$,
L. Moscardini$^{7,8}$\\
\\
$^1$ INAF - Osservatorio Astronomico di Trieste, Via G.B. Tiepolo 11,
I-34131 Trieste, Italy
(viel@oats.inaf.it)\\
$^2$ INFN/National Institute for Nuclear Physics, Via Valerio 2,
I-34127 Trieste, Italy\\
$^{3}$ Dipartimento di Fisica, Universit\`a di Roma TRE,
via della Vasca Navale 84, I-00146, Roma, Italy
\\
$^4$ Max-Planck Institut fuer Astrophysik,
Karl-Schwarzschild Strasse 1, D-85748 Garching, Germany\\
$^{5}$ Dipartimento di Fisica, Universit\`a di Padova,
via Marzolo 8, I-35131, Padova, Italy
\\
$^{6}$ INFN, Sezione di Padova,
via Marzolo 8, I-35131, Padova, Italy\\
$^{7}$ Dipartimento di Astronomia, Universit\`a di Bologna,
via Ranzani 1, I-40127 Bologna, Italy\\
$^{8}$ INFN, Sezione di Bologna, viale Berti Pichat 6/2,
I-40127 Bologna, Italy\\
\\}

\begin{document}
\maketitle
\begin{abstract}
We present results from the first high-resolution hydrodynamical
simulations of non-Gaussian cosmological models. We focus on the
statistical properties of the transmitted \lya flux in the high
redshift intergalactic medium. Imprints of non-Gaussianity are present
and are larger at high redshifts. Differences larger than 20\% at
$z>3$ in the flux probability distribution function for high
transmissivity regions (voids) are expected for values of the non
linearity parameter $f_{\rm NL}=\pm 100$ when compared to a standard
$\Lambda$CDM cosmology with $f_{\rm NL}=0$.  We investigate also the
one-dimensional flux bispectrum: at the largest scales (corresponding
to tens of Mpc) we expect deviations in the flux bispectrum up to 20\%
at $z\sim 4$ (for $f_{\rm NL}=\pm 100$), significantly larger than
deviations of $\sim 3\%$ in the flux power spectrum. We briefly
discuss possible systematic errors that can contaminate the signal.
Although challenging, a detection of non-Gaussianities in the
interesting regime of scales and redshifts probed by the \lya forest,
could be possible with future data sets.
\end{abstract}


\section{Introduction}

According to the standard gravitational instability picture
present-day cosmic structures have evolved from tiny initial
fluctuations in the mass density field that obey Gaussian statistics.
However, departures from Gaussianity inevitably arise at some level
during the inflationary epoch.  The various mechanisms that produce
primordial non-Gaussianity during inflation have been thoroughly
investigated by \cite{bartolo04} (and references therein). A
convenient way of modeling non-Gaussianity is to include quadratic
correction in the Bardeen's gauge-invariant potential $\Phi$:
\begin{equation}
\label{FNL}
\Phi = \Phi_{\rm L} + f_{\rm{\rm NL}} \left(\Phi_{\rm L}^2 - 
\langle\Phi_{\rm L}^2\rangle \right) \; ,
\end{equation}
where $\Phi_{\rm L}$ represents a Gaussian random field and the
dimensionless parameter $f_{\rm{\rm NL}}$ quantifies the amplitude of
the corrections to the curvature perturbations. The above definition
in which the term $-f_{\rm{\rm NL}}\,\langle\Phi_{\rm L}^2\rangle$ is small
guarantees that $<\Phi>=<\Phi_{\rm L}>=0$. Although the quadratic model
quantifies the level of primordial non-Gaussianity predicted by a
large number of scenarios for the generation of the initial seeds for
structure formation (including standard single-field and multi-field
inflation, the curvaton and the inhomogeneous reheating scenarios),
one should keep in mind that there are different ways for a density
field to be non-Gaussian (NG) and that different observational tests
capable of going beyond second order statistics should be used to
fully characterize the nature of non-Gaussianity.

To date, the strongest observational constraint for NG models are
provided by the recent analysis of the WMAP 5-year temperature
fluctuation maps \citep{komatsu08} according to which $-9 < f_{\rm NL}
< 111$ at the 95 \% confidence level in the local model.  The large
scale structure (LSS) provides alternative observational constraints
which are, in principle, more stringent than the cosmic microwave
background (CMB) since they carry information on the 3D primordial
fluctuation fields, rather than on a 2D temperature map.  Moreover, if
the level of primordial non-Gaussianity depends on scale then CMB and
LSS provide independent constraints since they probe different scales.
For this reason, the WMAP 5-year limits on $f_{\rm{\rm NL}}$ need not to
apply on the smaller scales probed by the LSS and the NG models that
we consider in this work, which have $|f_{\rm NL}|$ as large as 200,
are thus not in conflict with the CMB on the scales which are relevant
for our analysis.

A very promising way to constrain departures from Gaussianity is to
measure the various properties of massive virialized structures like
their abundance \citep{matarrese00, VJKM01, loverde08}, clustering and
their biasing \citep{GW86, MLB86,
  matarrese08,DDHS07,carbone08,seljak08,matarrese08}.  Indeed, the
best constraints on non-Gaussianity from the LSS have been obtained by
\cite{slosar08} including the observed scale-dependent bias of the
spectroscopic sample SDSS luminous red galaxies and the photometric
quasar sample. The resulting limits of $-29 < f_{\rm NL} < 70$ (95 \%
confidence level) are remarkably close to those obtained from the CMB
analysis alone and, according to \cite{seljak08}, could be further
improved by looking for scale dependency in the {\it relative} biasing
of two different population of objects.  Alternatively, one can
consider the topology of the mass density field \citep{matsubara03},
and higher-order clustering statistics like the bispectrum
\citep{hikage06}.  The ability of these techniques to detect the
imprint of the primordial non-Gaussianity on the LSS has been tested
with N-body experiments
\citep{messina90,moscardini91,weinbergcole92,mathis04,kang07,grossi07,DDHS07,hikage08}.
N-body simulations are of paramount importance in the study of NG
models, since one needs to disentangle primordial non-Gaussianity from
late non-Gaussianity induced by the non-linear growth of density
perturbations that can only be properly accounted for by numerical
experiments.

Recently, \cite{grossi07,grossi08} have carried out cosmological
N-body simulations of NG models to study the evolution of the
probability distribution function (PDF) of the density
fluctuations. They found that the imprint of primordial
non-Gaussianity, which is evident in the negative tail of the PDF at
high redshifts, is preserved throughout the subsequent evolution and
out to the present epoch.  This result suggests that void statistics
may be a promising effective tool for detecting primordial
non-Gaussianity (\cite{kamionkowski08,song08}) and that can be applied to different types of
observations over a large range of cosmic epochs.  Taking advantage of
the recent theoretical efforts for standardizing the appropriate
statistical tools \citep{colberg08} one could apply void-finding
algorithms to quantify the properties of the underdense regions
observed in the spatial distribution of galaxies.  Unfortunately,
current galaxy redshift surveys are probably too small for void-based
statistics to appreciate deviations from the Gaussian case at the
level required. The situation will change in a not too distant future,
when next generation all-sky surveys like ADEPT or EUCLID will allow
to measure the position of $\sim 5\times 10^7$ galaxies over a large
range of redshifts out to $z=2$. Alternatively, one can analyze
high-resolution spectra of distant quasars to characterize the
properties of the underlying mass density field at $z>3$
(e.g. \cite{viel03,viel04,vielbispect,lesg}).  In particular, since we
expect that underdense regions are characterized by a low neutral
hydrogen (HI) abundance, one can infer the presence of voids and
quantify their statistical properties from voids in the transmitted
flux, defined as the connected regions in the spectral flux
distribution above the mean flux level.  The connection between voids
and spectral regions characterized by negligible HI absorption has been
recently studied by \cite{viel08voids} using hydrodynamical
simulations where a link at $z\sim 2$ between the flux and matter
properties is provided.

In this work we perform, for the first time, high-resolution
hydrodynamical simulations of NG models to check whether one can use
the intergalactic medium (see \cite{meiksin07} for a recent review) to
detect non-Gaussian features in the \lya flux statistics like the PDF,
flux power and the bispectrum.  The layout of the paper is as follows.
In Section~ 2 we describe the hydrodynamical simulations and we show
an example of simulated \lya quasar (QSO) spectrum. In Section~ 3 we
present the results of the various flux statistics.  In Section~ 4 we
address the role of systematic and statistical errors that could
contaminate the NG signal. We conclude in Section~ 5.

\begin{figure*}
\begin{center}
\includegraphics[width=19cm]{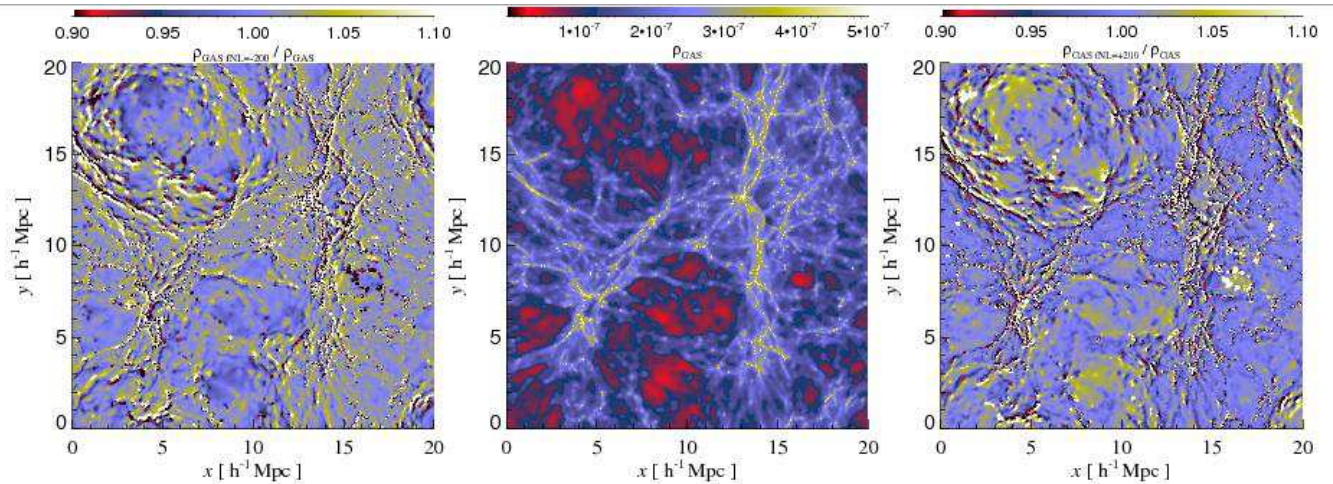}
\end{center}
\caption{Gas (projected) density distribution in the hydrodynamical
  simulations at $z=3$ in the Gaussian case (middle panel) and
  residuals in the gas distribution for non-Gaussian models with
  $f_{\rm NL}=-200$ (left panel) and $f_{\rm NL}=+200$ (right panel). The
  simulated box has a linear size of 20 comoving Mpc$/h$ and the
  thickness of the slice shown is 1.5 comoving Mpc$/h$. The cosmic web
  that gives rise to \lya absorption is visible and the tendency for
  regions below the mean density to be underdense (overdense) in
  models with negative (positive) values of $f_{\rm NL}$ is clear.}
\label{fig_slice}
\end{figure*}

\section{Non-Gaussian hydrodynamical simulations}
We rely on simulations run with the parallel hydrodynamical (TreeSPH)
code {\small {GADGET-2}} based on the conservative
`entropy-formulation' of SPH \citep{springel}.  They consist of a
cosmological volume with periodic boundary conditions filled with an
equal number of dark matter and gas particles.  Radiative cooling and
heating processes were followed for a primordial mix of hydrogen and
helium.  We assumed a mean Ultraviolet Background similar to that
propesed by \cite{haardt1996} produced by quasars and galaxies as
given by with helium heating rates multiplied by a factor 3.3 in order
to better fit observational constraints on the temperature evolution
of the IGM (e.g. \cite{schaye00,ricotti00}).  This background gives
naturally a hydrogen ionization rate $\Gamma_{-12}\sim 1$ at the
redshifts of interest here (e.g. \cite{bolt05,fg08}).  The star
formation criterion is a very simple one that converts in
collisionless stars all the gas particles whose temperature falls
below $10^5$ K and whose density contrast is larger than 1000 (it has
been shown that the star formation criterion has a negligible impact
on flux statistics).  More details can be found in \citep{viel04}.

The cosmological reference model corresponds to a `fiducial'
$\Lambda$CDM Universe with parameters, at $z=0$, $\Omega_{\rm m
}=0.26,\ \Omega_{\rm \Lambda}=0.74,\ \Omega_{\rm b }=0.0463$, $n_{\rm
  s}=0.95$, and $H_0 = 72$ km s$^{-1}$ Mpc$^{-1}$ and $\sigma_8=0.85$
(the B2 series of \cite{viel04}). We have used $2\times 384^3$ dark
matter and gas particles in a $60\ h^{-1}$ comoving Mpc box for the
flux power and bispectrum, to better sample the large scales. For the
flux probability distribution function we relied instead on $2\times
256^3$ dark matter and gas particles in a $20\ h^{-1}$ comoving Mpc,
since below and around $z=3$ this seems to be the appropriate
resolution the get numerical convergence.  The gravitational softening
was set to 2.5 and 5 $h^{-1}$ kpc in comoving units for all particles
for the 20 and 60 comoving Mpc/$h$ boxes, respectively. The mass per
gas particle is $6.12\times10^6 $M$_{\odot}/h$ for the small boxes and
$4.9\times10^7 $M$_{\odot}/h$ for the large boxes, while the high
resolution run for the small box has a mass per gas particle of
$1.8\times10^6 $M$_{\odot}/h$ (this refers to a (20,384) simulation
that was performed in order to check for numerical convergence of the
flux PDF). In the following, the different simulations will be
indicated by two numbers, $(N_1,N_2)$: $N_1$ is the size of the box in
comoving Mpc$/h$ and $N_2$ is the cubic root of the total number of
gas particles in the simulation.  NG are produced in the initial
conditions at $z=99$ using the same method as in \cite{grossi07} that
we briefly summarize here. Initial NG conditions are generated without
modifying the linear matter power spectrum using the Zel'dovich
approximation: a Gaussian gravitational potential is generated in
Fourier space from a power-law power spectrum of the form $P(k)\propto
k^{-3}$ and inverse Fourier transformed in real space to produce
$\phi_{\rm L}$. The final $\Phi$ is obtained using eq. (1).  Finally,
back in Fourier space, we modulate the power-law spectrum using the
transfer functions of the $\Lambda$CDM model.

We also run some other simulations at higher resolutions to check for
numerical convergence.  In particular we have performed a $(20,384)$
simulation run to analyse the flux PDF. For the $(20,256)$ models the
flux probability distribution function has numerically converged only below
$z=3$ (see \cite{bolton07}). However, since our results will
always be quoted in comparison with the $f_{\rm NL}=0$ case (i.e. as a
ratio of two different quantities) we expect the resolution errors to
be unimportant (i.e. we assume the same resolution corrections should
be applied to all the models, even though this assumption should be
explicitly checked).

A projected density slice of the gas (IGM) distribution for the
(20,256) simulation of thickness 1.5 comoving Mpc/$h$ is shown in
Figure \ref{fig_slice}. We focus on this simulation because at $z=3$
the flux probability distribution function has numerically
converged. In the middle panel we plot the gas
density in the Gaussian case, while residuals in the two models with
$f_{\rm NL}=-200$ and f$_{\rm NL}=+200$ are shown in the left and right panel,
respectively. On average regions of the cosmic web below the mean
density appear to be $\sim 10\%$ less (more) dense in the negative
(positive) $f_{\rm NL}$ case. This trend is apparent not only near the
centre of these regions but also in the matter surrounding them (see
for example the void at $(x=17; y=8)$ comoving Mpc$/h$). The same
qualitative behavior can be observed in the distribution of the dark
matter particles (see Figure 2 of \cite{grossi08}).

\begin{figure*}
\begin{center}
\includegraphics[width=19cm]{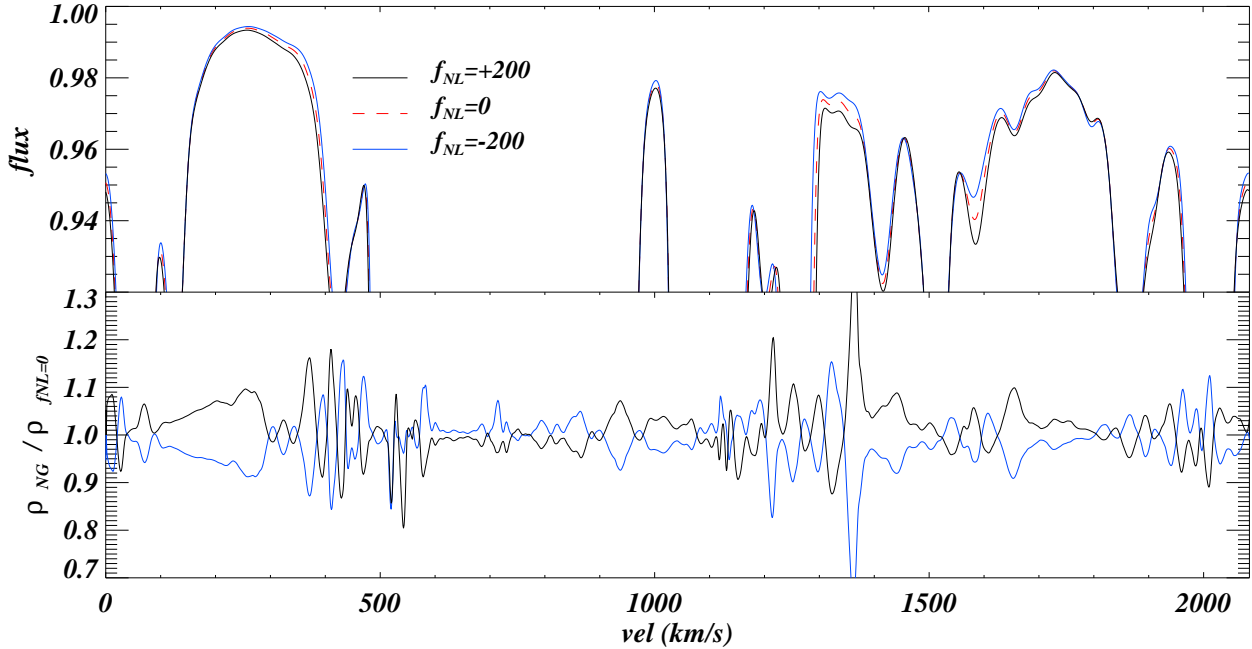}
\end{center}
\caption{Simulated noiseless \lya flux (only regions at high
  transmissivity are shown) at $z=3$ in the upper panel. Models with
  positive and negative values of $f_{\rm NL}$ bracket the Gaussian
  case with $f_{\rm NL}=0$. Although difficult to see on a
  pixel-by-pixel basis, differences among the models can be
  appreciated in a statistical sense when many spectra are considered
  (see text).  The bottom panel shows the ratio of non-Gaussian models
  to the Gaussian one for the one-dimensional gas density along the
  line-of-sight (in real space). On average differences of the order
  10\% are present.}
\label{fig_los}
\end{figure*}

In the NG models considered here the growth of structures in terms of
density PDF is different. As discussed in Grossi et al. (2008), the
maps of residuals in the non-Gaussian cases reflect the differences in
the primordial PDF of the mass overdensity.  As shown in Figures 1 and
5 of Grossi et al.  (2008) the mass PDF is skewed towards positive
(negative) overdensities in the non-Gaussian models with positive
(negative) $f_{\rm NL}$ values, compared to the Gaussian case.  As a
consequence, since the gas traces well the underlying mass
distribution at these redshifts, voids look emptier in the $f_{\rm
  NL}=-200$ case (map on the left) while denser environments like
filaments and knots look more prominent in the $f_{\rm NL}=+200$ case
(map on the right) with respect to the Gaussian case. These
differences in the tails of the density PDF also impact on the
filaments at around the mean density that surround the voids. In fact
the size of the voids is slightly different in the negative and
positive NG models: for negative $f_{\rm NL}$ values the emptier voids
grow in size faster than for the Gaussian case and even faster than
for positive $f_{\rm NL}$ values, displacing the filaments at around
the mean densities at different positions in the three cases and giving
rise to the filamentary pattern of residuals of the panels.

To perform our analysis we have extracted several mock QSO absorption
spectra from the simulation box.  All spectra are drawn in redshift
space taking into account the effect of the IGM peculiar velocities
along the line-of-sight $v_{\rm pec,\parallel}$. Basically, the simulated
flux at the redshift-space coordinate $u$ (in km/s) is
$F(u)=\exp[-\tau(u)]$ with: \be \tau(u)={\sigma_{0,\alpha} ~c\over H(z)}
\int_{-\infty}^{\infty} dx\, n_{\rm HI}(x) ~{\cal
  G}\left[u-x-v_{pec,\parallel}^{\rm
    IGM}(x),\,b(x)\right]dx \label{eq1} \;, \ee where
$\sigma_{0,\alpha} = 4.45 \times 10^{-18}$ cm$^2$ is the hydrogen \lya
cross-section, $H(z)$ is the Hubble constant at redshift $z$, $x$ is
the real-space coordinate (in km s$^{-1}$), $b=(2k_BT/mc^2)^{1/2}$ is
the velocity dispersion in units of $c$, ${\cal G}=(\sqrt{\pi}
b)^{-1}\exp[-(u-y-v_{\rm pec,\parallel}^{\rm IGM}(y))^2/b^2]$ is the
Gaussian profile that well approximates the Voigt profile in the
regime considered here. The neutral hydrogen density in real-space,
that enters the equation above, could be related to the underlying gas
density by the following expression (e.g. \cite{huignedin97,schaye}):
\begin{eqnarray} 
n_{\rm HI}({\bf x}, z) &  \approx
10^{-5} ~{\overline n}_{\rm IGM}(z) \left({\Omega_{0b} h^2 \over
  0.019}\right) \left({\Gamma_{-12} \over 0.5}\right)^{-1} \times \nonumber \\
&  \left(T({\bf x},z) \over 10^4 {\rm K} \right)^{-0.7} \left({1+z \over
  4}\right)^3 \left(1 + \delta_{\rm IGM}({\bf x},z) \right)^2 \;,  
\end{eqnarray}
with $\Gamma_{-12}$ is the hydrogen photoionization rate in units of
$s^{-1}$, $T$ is the IGM temperature, and $\overline{n}_{\rm IGM}(z)$
is the mean IGM density at that redshift. However, this equation is
not explicitly used since the neutral hydrogen fraction is computed
self-consistently for each gas particles during the simulation run.
The integral in eq. (\ref{eq1}) to obtain the \lya optical depth along
each simulated line-of-sight is thus performed using the relevant
hydrodynamical quantities from the numerical simulations: $\delta_{\rm
  IGM}, T, v_{\rm pec}, n_{\rm HI}$. More details on how to extract a
mock QSO spectrum from an hydrodynamical simulation using the SPH
formalism can be found in \cite{theuns98}.

An example of line-of-sight is shown in the top panel of Figure
\ref{fig_los}, while the bottom panel shows the ratio of the gas
density along the line-of-sight of NG and Gaussian models (in real
space) In the following we will focus on high transmissivity in which
the transmitted flux is close to unity (upper panel). Three QSO
spectra are shown with different line styles and correspond to the
Gaussian case (dashed, red line), and to $f_{\rm NL}=\pm200$ (solid black
and solid blue, respectively).  The transmitted flux (no noise is
added in this case) is almost identical for the two NG models in
magnitude but not in sign, as expected.  One can better appreciate the
differences among the models by looking at the gas density (bottom
panel).  On average differences are of the order 10\%, even if in some
cases they can rise above 30-40\%.  The fact that the corresponding
variations in the flux are comparatively smaller (usually less than
few percent) is somehow expected, since differences in the gas density
are exponentially suppressed by the non-linear transformation between
flux and matter (and by other non-linear effects as well).  However,
despite their small amplitude, the differences in the
transmitted flux are large enough to be appreciated through
appropriate statistical analyses of many independent lines-of-sight,
as we will see in the following sections.  Global statistics will be
usually shown for samples of 1000 lines-of-sight extracted along
random directions within the simulated volume.

\begin{figure*}
\begin{center}
\includegraphics[width=18cm, height=7cm]{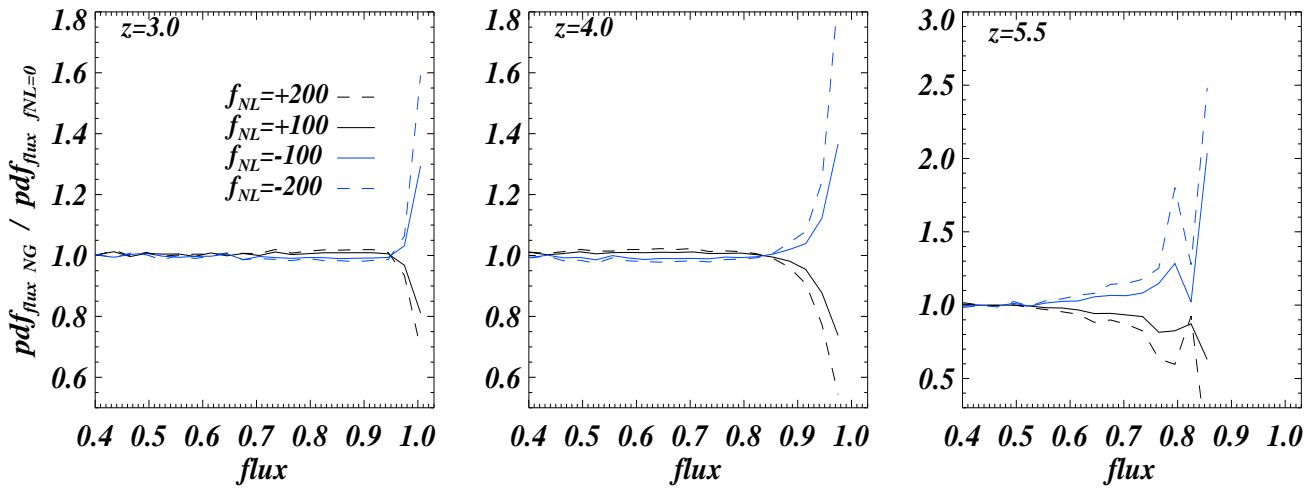}
\end{center}
\caption{Ratio between the simulated flux probability distribution
  function of four different $(20,256)$ models with
  $f_{\rm NL}=-200,-100,+100,+200$ and $f_{\rm NL}=0$, represented by the blue
  dashed, blue continuous, black continuous and black dashed lines,
  respectively. Results are shown at $z=3,4,5.5$ in the left, middle and
  right panels. (Note the different scale in the right panel). }
\label{fig1}
\end{figure*}

\section{Results}

\subsection{The \lya flux probability distribution function}

In Figure \ref{fig1} we show results for the flux probability
distribution function at $z=3,4 \ {\rm and} \ 5.5$ using the (20,256)
simulations, in the left, middle and right panel, respectively. The
mock QSO spectra have been normalized to reproduce the same (observed)
mean flux level. The scaling factor is usually different by less
than 2\% to the standard non-Gaussian case (more precisely the
differences are below 1\% at z=3, around 1\% at z=4.0 and around 2\%
at z=5.5 between the Gaussian and the $f_{\rm NL}=\pm 200$ cases).
Differences between the non-Gaussian case and the Gaussian one are
appreciable only in regions of high transmissivity (flux $\sim$ 1),
that are typically associated to connected regions below mean density
(voids) in the matter distribution.  At $z=3$ (left panel) the
differences can be of the order of 20\% (40\%) for models with $f_{\rm
  NL}=\pm\, 100 (\pm\, 200)$. Models with negative (positive) values
of $f_{\rm NL}$ produce less (more) absorption. This reflects the fact
that voids in models with negative $f_{\rm NL}$ are emptier of neutral
hydrogen than in the Gaussian case. The opposite holds true for models
with positive $f_{\rm NL}$.  This is analogous to the effect discussed
by \cite{grossi08} on the dark matter density field and characterized
in terms of the probability distribution function of density
fluctuations. In that case for negative values of $f_{\rm NL}$ the low
density tail of the dark matter density PDF is more prominent. In our
case what is more prominent is the high flux tail of the \lya flux
PDF.  The amplitude of the effect increases with the redshift.  At
$z=4$ (middle panel) differences w.r.t. the Gaussian case are as large
as 30\,\%-60\,\% (for $f_{\rm NL}=\pm 100$ and $\pm 200$,
respectively) and at $z=5.5$ (right panel) the differences are of the
order of $\sim 100\,\%-150\%$ (for $f_{\rm NL}=\pm 100$ and $\pm 200$,
respectively). Note that in the latter ($z=5.5$) case we have used a
different scale for the $y-$axis.

From an observational viewpoint, it should be noted that the \lya flux
PDF has been measured with great accuracy using high-resolution
spectra taking into account the metal contaminations and continuum
fitting errors at $z=2.07,2.54,2.94$ by \cite{tkim}.  On the contrary,
continuum fitting errors and the metal contaminations are somewhat
harder to estimate in the measurements at higher redshifts ($z=4.5$
and $z=5.5$) by \cite{Becker:2006qj}. We will come back to this point
in Section 4.

\subsection{The \lya flux void distribution function}

A different, although not completely unrelated statistics is
represented by the probability distribution function of the voids of
given comoving size $R$. Searching for voids in the \lya forest of
observed QSO spectra has a long dating history (see for example
\cite{carswellrees1987,crotts,duncan,ostriker,dobrzycki,rauch92}) but
in this paper we focus on the impact of non-Gaussianities on their
statistical properties.

\begin{figure}
\begin{center}
\includegraphics[width=8cm]{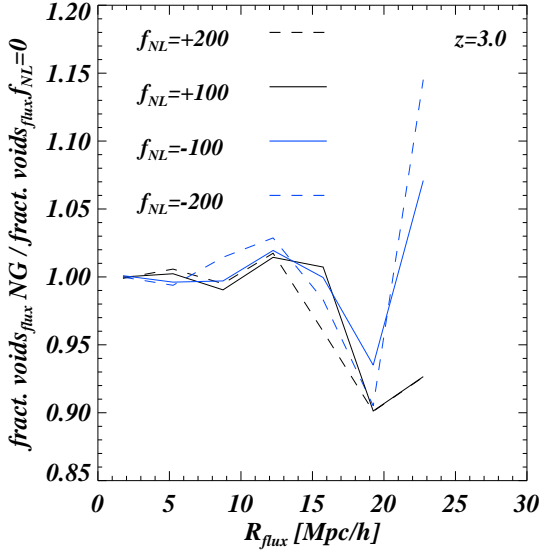}
\end{center}
\caption{Ratio of the fraction of voids in the flux distribution
  between NG models and the corresponding Gaussian one at $z=3$ as a
  function of void size in comoving Mpc$/h$. The models with $f_{\rm
    NL}=-200,-100,+100,+200$ are represented by the blue dashed, blue
  continuous, black continuous and black dashed lines, respectively.}
\label{fig5}
\end{figure}

We define flux voids as in \cite{viel08voids}: connected
one-dimensional regions along the QSO spectrum whose transmitted flux
is above the mean flux level at that redshift.  In Fig.~\ref{fig5} we
show the ratio between the probability distribution functions of the
NG and Gaussian cases at $z=3.0$. For this plot we have used the
(60,384) that have the largest box size. Although the size of the
largest voids (R$\sim 20$ comoving Mpc$/h$) is comparable to that of
the box the corresponding differences in the probability distribution
are rather mild and of the order of 10-15\%.

The differences are as expected for voids of sizes larger than 20
comving Mpc$/h$ (while for smaller ones the differences are
negligible): negative values of $f_{\rm NL}$ result in voids that are
emptier compared to the standard Gaussian case and thereby the typical
sizes could be larger; while the opposite trend can be seen for the
positive values of $f_{\rm NL}$.  The effect, even for $f_{\rm NL}=\pm
200$, is however somewhat smaller than the effects that can be induced
by changing other cosmological or astrophysical parameters (see the
relevant plots in \cite{viel08voids}).  Furthermore, the uncertainty
in the mean flux level at $z=3$, which enters the definition of a void
in the flux, produces an effect that is still larger than the NG
signal sought.

\subsection{The \lya flux power spectrum}
\begin{figure*}
\begin{center}
\includegraphics[width=18cm, height=7cm]{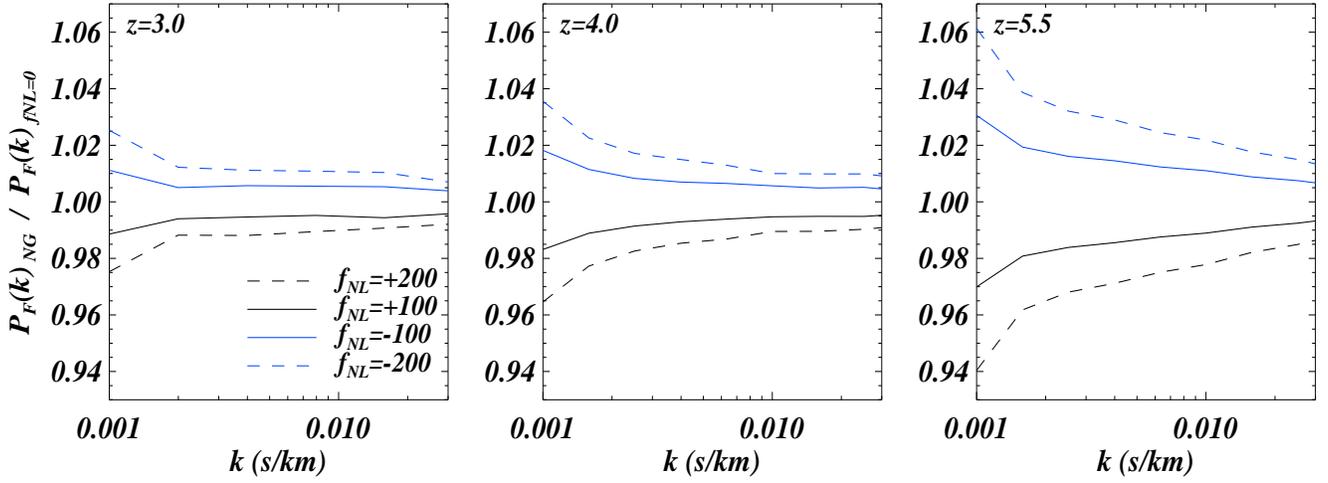}
\end{center}
\caption{Ratio between the simulated 1D flux power spectrum of four
  different $(60,384)$ models with $f_{\rm NL}=-200,-100,+100,+200$ and
  $f_{\rm NL}=0$, represented by the blue dashed, blue continuous, black
  continuous and black dashed lines, respectively. Results are shown
  at at $z=3,4 \ {\rm and} \ 5.5$ in the left, middle and right
  panels, respectively.}
\label{fig2}
\end{figure*}

Primordial non-Gaussianity affects the evolution of density
perturbation, particularly at the epochs and scales in which they
enter the non-linear regime.  Deviations from the Gaussian case are
larger at high redshift, since at late times nonlinear dynamical
effects become dominant.  However, the contribution non-Gaussianity
implied by $f_{\rm NL}=\pm100$ is always within a few percent of the total
gravitational potential and should not appreciably affect the linear
matter power spectrum.  Consequently, we also expect the effect on the
1D flux power spectrum to be small.

To quantify the effect we have plotted in Figure \ref{fig2} the 1D
flux power spectrum for the Gaussian and non-Gaussian cases at $z=3,4,
\ {\rm and} \ 5.5$, in the left, middle and right panels respectively.
Even in this case the QSO spectra have been normalized to reproduce
the same mean flux.  Differences to the Gaussian case are of the order
of 2\%, 3\% and 5\% at the redshifts considered here and manifest
themselves as an overall plateau with slightly more power at the
largest scales (a factor two larger than at the smallest scales
probed). As expected, the effect of primordial non-Gaussianity on the
flux power spectrum is small and the effect decreases with time.

In principle this effect on the flux power is degenerate only with a
change in the mean flux level (see for example Figure 3 of
\cite{vielhaehnelt06} or Figure 13 of \cite{mcdonald05}): this means
that other changes in cosmological parameters and/or astrophysics
produce a different $k$-dependent change in the flux power than the
one produced by non-Gaussianities. However, the magnitude of this
effect is quite small and probably not detectable with present data
sets.

\subsection{The \lya flux bispectrum}
Unlike the power spectrum, the bispectrum on large scales is sensitive
to the statistical properties of primordial fluctuations like a
primordial non-Gaussianity \citep{fry94,verde02,sefusatti08}.
Therefore, the 1D flux bispectrum looks like a very promising
statistics to search for non-Gaussianities in the IGM.  The \lya flux
bispectrum has been calculated for the first time using
high-resolution QSO spectra by \cite{vielbispect}. Here, we use the
same definition i.e. the real part of the three point function in
$k-$space $, D_F= {\rm Re}(\delta_F (k_1)\, \delta_F (k_2) \, \delta_F
(k_3))$, for closed triangles $ k_1 + k_2 + k_3 = 0$.  $\delta_F(k)$
is the Fourier transform of $\delta F$.  $D_F$ is related to the {\it
  bispectrum of the flux} $B_F(k_1,k_2,k_3)$ \be \langle D_F \rangle =
2\,\pi \,B_F(k_1, k_2, k_3)\,\delta^D (k_1 + k_2 + k_3) \; .
\label{eq2}
\ee $\delta^D(k)$ is the one-dimensional Dirac delta function and
$\langle \cdot \rangle$ indicates the ensemble average.  Since we
compute the one-dimensional bispectrum our triangles are degenerate
and we choose two configurations: $i)$ the flattened configurations
for which $k_1=k_2$ and $k_3 = -2\,k_1$; $ii)$ the squeezed
configuration for which $k_1=k-k_{\rm min}$, $k_2=-k-k_{\rm min}$ and $k_3 =
2\,k_{\rm min}$, with $k_{\rm min}=2\pi/L$ (L the linear size of the box in
km/s). In the following we will always show the flux bispectrum as a
function of the wavenumber $k=k_1$.  In \cite{vielbispect} a numerical
calculation of the flux bispectrum was compared to analytical
estimates obtained through an expansion at second order of the
fluctuating Gunn-Peterson approximation (\cite{FGPA}): while the
overall amplitude of the bispectrum was not matched by the theory, the
shape, at least at large scales, was well reproduced. However, the
theoretical expression for the flux bispectrum contained only the
gravitational terms. Here we extend this work by computing the flux
bispectrum for NG Gaussian models using the numerical
hydrodynamical simulations performed.

\begin{figure*}
\begin{center}
\includegraphics[width=16cm]{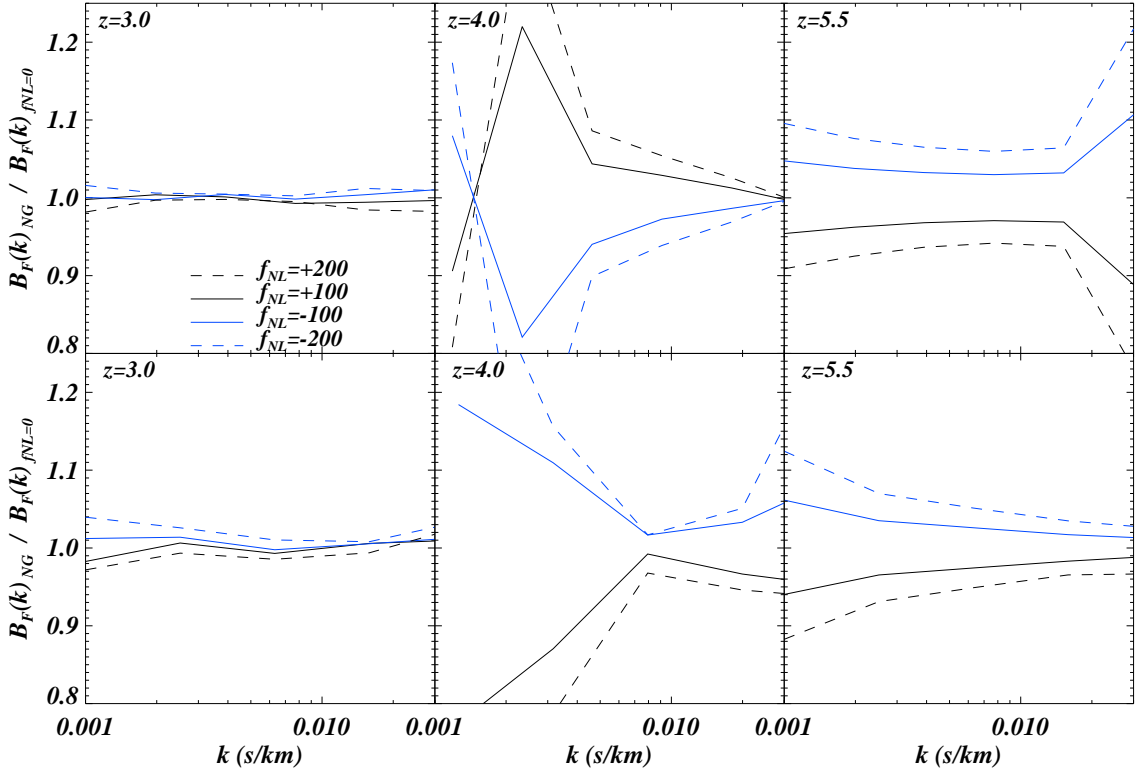}
\end{center}
\caption{Ratio between the simulated 1D flux bispectrum in the
  flattened $(k,k,-2k)$ (bottom row) and squeezed configuration (top
  row) $(k-k_{min},-k-k_{min},2k_{min})$ with $k_{min}=2\pi/L$ (and
  with $L$ the linear size of the box in km/s) of four different $(60,384)$
  models with $f_{\rm NL}=-200,-100,+100,+200$ and $f_{\rm NL}=0$, represented
  by the blue dashed, blue continuous, black continuous and black
  dashed lines, respectively. Results are shown at $z=3,4,5.5$ in the
  left, middle and right panel, respectively.}
\label{fig3}
\end{figure*}

In Figure \ref{fig3} we plot our findings in terms of ratios between
the Gaussian and non-Gaussian models in the squeezed (top panels) and
flattened (bottom panels) configurations. Due to the intrinsic noisy
nature of the bispectrum, we have binned the values in $k-$space, in
the same way as the flux power of the previous subsection.

One can see that while at $z=3$ the differences are very small and
usually less than 3-4\%, they become much larger and of the order of
30-40\% at $z=4$. At $z=5.5$ the differences become again smaller and
with different wavenumber dependence. It is possible to interpret this
trend in the framework of the second order perturbation theory as done
in \cite{vielbispect}: the overall amplitude and shape of the flux
bispectrum could not be smooth and strongly depend (in a non-trivial
way) on the redshift evolution of the coefficients that describe the
evolution of the mean flux level and of the IGM temperature-density.

\section{Discussion}

\begin{figure*}
\begin{center}
\includegraphics[width=18cm, height=7cm]{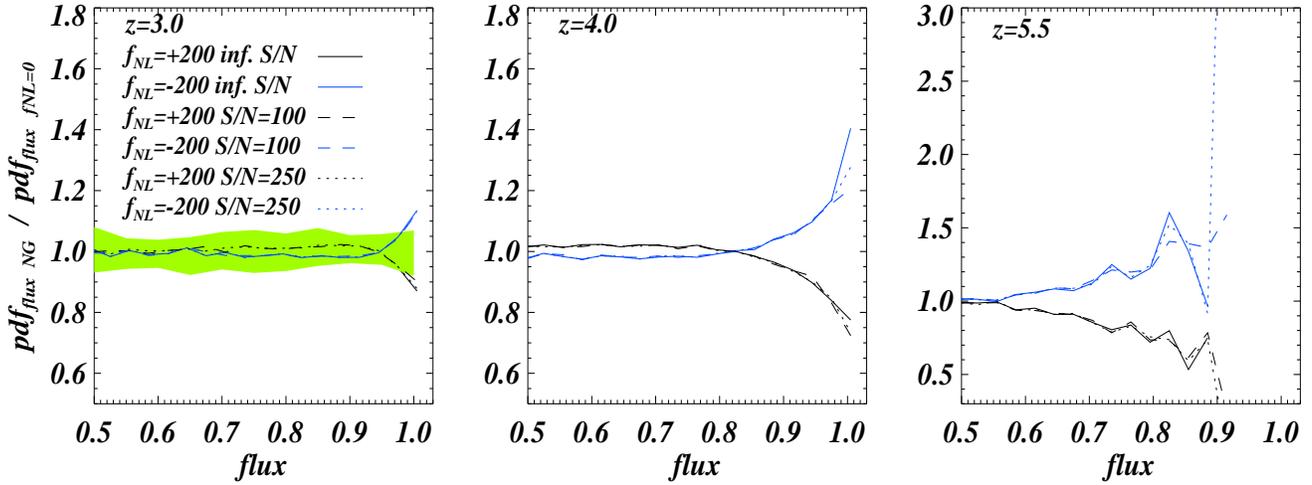}
\end{center}
\caption{Ratio between the simulated flux probability distribution
  function of two different $(20,256)$ models with $f_{\rm
    NL}=-200,+200$ and the gaussian model with $f_{\rm NL}=0$, represented by the blue and
  black lines, respectively. The continuous lines are for an infinite
  signal-to-noise ratio, while dashed and dotted are for
  signal-to-noise ratios of 100 and 250.  Results are shown at
  $z=3,4,5$ in the left, middle and right panels. The continuum errors
  are modelled as described in the text and are of the order 2\% at
  $z=3$ and 10\% at $z=4,5.5$.  In the left panel we report as a
  filled area the statistical error bars of the flux PDF at $z=2.94$,
  as found in \citep{tkim}.}
\label{fig_syst}
\end{figure*}

Among the different flux statistics that we have explored, the flux
PDF seems the most promising in order to detect primordial
non-Gaussiantity.  However, to assess whether such information can
actually be extracted from the real datasets one needs to compare the
expected signal with the amplitude of the known errors.  The present
statistical uncertainties in the flux PDF at $z<3$ derived from
high-resolution high signal-to-noise spectra is below 4-5\%.  This
number was derived using jack-knife estimators from a suite of
high-resolution high signal-to-noise ($> 50$ and usually around 100)
QSO spectra by \cite{tkim}. This is smaller than the effect we are
seeking and possibly the NG signature is degenerate with other effects
such as a change in the temperature evolution of the IGM.

In Figure \ref{fig_syst} we show in a quantitative way the effect of
the observational errors on the flux PDF, at $z=3,4 \ {\rm and}
\ 5.5$.  We use a realistic (observed) array of signal-to-noise values
taken from \cite{tkim} at $z=2.94$.  The signal-to-noise depends on
the transmitted flux, and the average signal-to-noise value for the
noise array taken turned out to be $\sim 100$. 

The various curves in these plots are the same as in Fig.~\ref{fig1}
for the $f_{\rm NL}=\pm 200$ cases only. The dashed refers to the
realistic errors of \cite{tkim} corresponding to an average $S/N=100$.
We also plot the case of a more favorable case with $S/N=250$ (dotted
curve), while the infinite signal-to-noise error is represented by the
continuous line.  A second source of uncertainty is represented by
continuum fitting errors that we have modelled in a statistical way
that produces a $\pm 2\%$, $\pm 6\%$ and $\pm 10\%$ displacement of
the continuum level at $z=3$ and at $z=4$ and $z=5.5$, respectively.
These numbers have been derived by the estimates of \cite{tkim} and
\cite{Becker:2006qj} based on the analysis of high resolution and high
signal-to-noise QSO spectra. To account for these errors we have
adjusted the simulated continuum of the transmitted flux along every
line-of-sight by factor $1\pm G$, where $G$ is a number drawn from
Gaussian distributions with width 0.02, 0.06 and 0.1 at $z=3,4$ and
$z=5.5$.  This should provide a reasonable estimate of the continuum
fitting errors effects on the flux PDF as long as these errors are
Gaussian.

Of course, taking into account realistic signal-to-noise values and
the continuum fitting errors reduces the significance of the NG
signal. We find that for a signal-to-noise ratio of 100 (250) it is
reduced by ~ 40 \% (20\%) at $z=3$, in the same way at higher
redshifts, where the NG signal is higher, we find similar values. The
continuum fitting errors are somewhat more important and reduce the
significance of the NG signal on the flux PDF for $f_{\rm NL}=\pm 200$ by
$\sim 40\%$ at $z=3$. However, adding the two sources of errors at the
same time as shown in Figure \ref{fig_syst} decreases the NG signal by
45 \%, 40\% and 80\% at $z=3,4,5.5$ for the $f_{\rm NL}=\pm 200$ cases,
respectively.

% The results are shown in Figure~\ref{fig_syst}
%for the usual three redshifts considered as blue ($f_{\rm NL}=-200$) and
%black ($f_{\rm NL}=+200$) diamonds.  By comparing these plots with those
%in Figure~\ref{fig1} we notice that continuum fitting errors reduce
%the significance of the NG signal on the flux PDF for $f_{\rm NL}=\pm 200$
%by $\sim 40\%$. 

%Note that this source of error is larger than the one
%introduced by the signal-to-noise and potentially could reduce the
%significativity of the NG signal at higher redshifts as well, where
%the continuum uncertainties are higher.  

The statistical errors estimated by \cite{tkim} are represented by the
shaded area in the leftmost panel and refers to $z=2.94$. Ideally, one
would like the NG signal to be larger than the statistical errors once
all the systematic errors have been taken into account. At $z=3$ we
are indeed in this case, but only marginally so. We find that the
effect of including continuum uncertainties at $z=3$ has the same
quantitative effect of dealing with a signal-to-noise ratio of 250
instead of an infinite one, and when these two errors are added
together the effect on the flux PDF is of the same order of the NG
signal for $f_{\rm NL}=\pm 200$. At higher redshifts the situation becomes
slightly better.  However, despite the reduction of the NG signal, its
signature is still large enough to be detected, especially at $z>3$,
and a higher significance could be of course reached once all the
\lya flux statistics (PDF, flux power and bispectrum) will be fitted
at the same time.

We stress that our quantitative arguments do not include the possible
degeneracies on the flux PDF of NG with other cosmological and
astrophysical parameters as addressed in \cite{bolton07}. It is
however intriguing that a better fit to the PDF data presented there
at $F>0.8$ would require emptier voids and thus negative values of
$f_{\rm NL}$.

\section{Conclusions and perspectives}

In this work we have explored the possibility of constraining
primordial non Gaussianity through the statistical properties of \lya
forest QSO spectra at $z\ge 3$. For this purpose, and for the first
time, we have performed a suite of high resolution non-Gaussian
hydrodynamical simulations.  Although recent analyses have provided
convincing evidence that the most stringent constraints to primordial
non-Gaussianity will be likely provided by the large scale biasing
properties of rare, massive objects (e.g. \cite{slosar08}), the
analysis of the \lya forest has to be regarded as complementary since
it would probe non-Gaussianity on smaller scales and at intermediate
epochs between other LSS probes and the CMB.

The main results of this study can be summarized as follows:

{\em (i)} the differences between the Gaussian and non-Gaussian scenarios 
are more evident in regions of high flux transmissivity associated to 
low density environments in the gas distribution;

{\em (ii)} deviations from the Gaussian case are best seen in the high
flux tail of the 1D flux PDF:  differences are of
the order of 20-30 \% and $z=3$ and increase up to $\sim 100$ \% at
$z=5.5$;

{\em (iii)} differences in the void distribution function are
comparatively smaller, indicating that the PDF is a better statistics
to spot primordial non-Gaussianity;

{\em (iv)}  the 1D flux power spectrum is little affected by 
non-Gaussianity, as expected by the analogy with the 
matter power spectrum: the measured differences are of the order of 
a few per cent and increase at higher redshifts;

{\em (v)}  the flux bispectrum represents a much more powerful 
statistics and potentially could provide strong constraints;

{\em (vi)} the significance of the non-Gaussian signal is highly
reduced when one accounts for realistic signal-to-noise values in the
measured flux PDF and continuum fitting errors at high redshifts;
nevertheless, significant constraints on the non-Gaussianity can still
be extracted from the analysis of the high flux tail of the flux PDF.

The statistical error bars on the flux power as measured using the
SDSS data release 3 by \cite{mcdonald06} are usually in the range 3-10
\% (going from the small scales 0.01 s/km to the largest 0.001 s/km)
in the range $z=2-4$, so the NG signal in this case is smaller than
the statistical error (even though combining all the data points the
error will be of the order on the power spectrum amplitude will become
0.6\% and on its slope of $\pm$ 0.005).  The SDSS data release 3 is
based on a sample of 3035, increasing the number of observed QSO
spectra will reduce further the statistical error by a factor
$\sqrt{N_{QSO}}$ making the NG signature more evident, once the
degeneracies with all the other cosmological and astrophysical
parameters will be properly addressed.

Regarding the flux bispectrum, the present statistical error bars at
$z\sim 2$ are of the order 50\% \cite{vielbispect}, as derived from high-resolution spectra, a value that is
much larger than what is expected from a non-Gaussian signal at that
redshift, while this value is comparable to what could be seen at
$z\sim 4$. Even in this case, in order to study in a precise way
putative NG signatures in the flux bispectrum, more work is needed to
address numerical convergence of the flux bispectrum and to
incorporate the relevant physical processes that can affect its shape
and amplitude down to smaller scales than those probed by the flux
power.

The statistical error bars derived from present data sets of QSO spectra at
high resolution in the flux PDF function are usually below 5\% for high
transmissivity regions. This value is basically determined by the
signal-to-noise ratio of the spectra and, at least potentially, higher
signal-to-noise ratios can be achieved and beat down this statistical
error. This statistics seems promising due to the large number of QSO
spectra available and to the better understanding of
systematics. Among the possible systematics the most important is the
uncertainty due to the continuum fitting errors which however could
probably be significantly reduced at high redshifts with a better
understanding  and removal of the QSO continuum.

\section*{Acknowledgments.}
Numerical computations were done on the COSMOS supercomputer at DAMTP
and at High Performance Computer Cluster Darwin (HPCF) in Cambridge
(UK). COSMOS is a UK-CCC facility which is supported by HEFCE, PPARC
and Silicon Graphics/Cray Research. Part of the analysis was also
performed at CINECA (Italy). We thank Francesca Iannuzzi for help with
the initial conditions NG-generator code. We acknowledge support from
ASI/INAF under contracts: I/023/05/0 I/088/06/0 e I/016/07/0. We thank
the referee Tom Theuns for a useful referee report.

\bibliographystyle{mn2e} \bibliography{master2.bib}

\end{document}